\documentclass[12pt]{iopart}

\usepackage{iopams}  
\usepackage{color}
\usepackage{graphicx}
\begin{document}

\title{On the limit of spectral gap formation via Bragg's reflection}
\author{Lei Chang}
\address{School of Aeronautics and Astronautics, Sichuan University, Chengdu 610065, China}
\ead{leichang@scu.edu.cn}
\date{\today}

\begin{abstract}
Although it is well recognized that waves propagating in periodic media have forbidden gaps in the continuous spectrum, it is still yet unknown about the minimum number and depth of periodic modulations for clear spectral formation. This limit is explored by investigating shear Alfv\'{e}n waves propagating through magnetic mirror array in a low temperature plasma cylinder, referring to an existing experiment [Zhang et al., Phys. Plasmas 15, 012103 (2008)]. It is found that the bottom-edge ratio, a defined parameter characterizing how ``noisy" the formed spectral gap is, scales inversely with the square of the number and depth of magnetic mirrors, and they should be bigger than $7$ and $0.4$ respectively to form a clear spectral gap for the plasma conditions employed. This scaling is consistent with a pervious analysis [Kryuchkyan and Hatsagortsyan, Phys. Rev. Lett. 107, 053604 (2011)], and also applicable to other fields such as optics and semiconductors. Moreover, it shows that the center of spectral gap experiences a parabolic shape of descending frequency shift when the modulation depth is increased, which is believed to be a new discovery. 

\end{abstract}

\textbf{Keywords:} Bragg's reflection, number and depth, descending frequency shift, magnetic mirror array, shear Alfv\'{e}n wave

\textbf{PACS:} 52.35.Bj, 42.25.Fx

\maketitle

As first discussed by Strutt (Lord Rayleigh) in $1887$, waves that propagate in periodic media experience periodic modulation in their index of refraction and thereby have forbidden gaps in the continuous spectrum, namely spectral gaps.\cite{Strutt:1887aa} The underlying physics is the interference between forward and backward waves reflected from periodic structure, and the gap is best formed when the Bragg's condition is satisfied ($k=n\pi L$ with $k$ the wave number, $n$ an integer and $L$ the length of period).\cite{Kittel:1996aa} The spectral gap due to Bragg's reflection widely occurs in many fields including plasma physics when the confining magnetic field or plasma density has periodic structure.\cite{Chu:1992aa, DIppolito:1980aa, Dewar:1974aa, Betti:1992aa, Nakajima:1992aa, Kolesnichenko:2001aa, Chang:2013aa} Zhang \etal.\cite{Zhang:2008aa} observed Alfv\'{e}nic spectral gap in a low temperature plasma cylinder, by setting up a periodic array of magnetic mirrors along the plasma column. However, the gap is not clear and global due to limited number of magnetic mirrors.\cite{Chang:2018aa} Although this limitation may be compensated by increasing the modulation depth,\cite{Chang:2016aa} it is still unknown about the minimum number and depth of periodic modulations for clear spectral gap formation.

\begin{figure}[ht]
\begin{center}
\includegraphics[width=0.7\textwidth,angle=0]{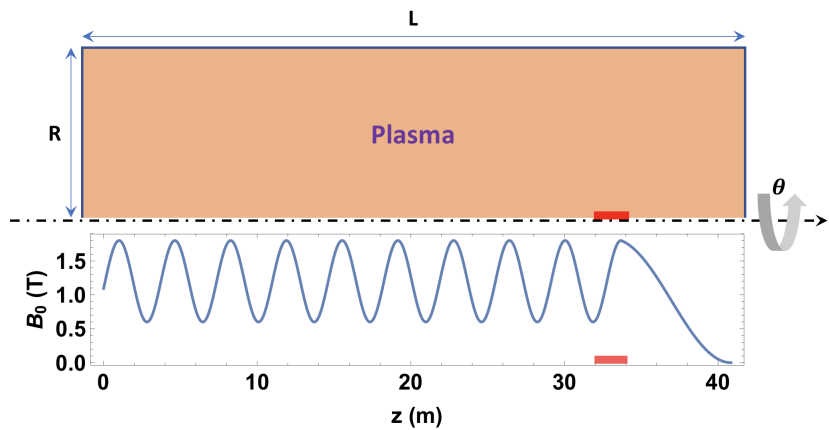}
\end{center}
\caption{A schematic of computational domain (above) and typical configuration of periodic magnetic field of $B(z)=1.2+0.6\cos(2\pi z/3.63)$~T with $9$ magnetic mirrors and beach of length $7.74$~m (below). The red bar denotes the blade antenna employed.}
\label{fg1}
\end{figure}

This letter explores the limit of spectral gap formation via Bragg's reflection with reference to Zhang \etal.'s experiment.\cite{Zhang:2008aa} Specifically, the effects of the number and depth of magnetic mirrors on Alfv\'{e}nic spectral gap are investigated in detail through full wave computations. The employed code is an ElectroMagnetic Solver (EMS),\cite{Chen:2006aa} based on Maxwell's equations and cold-plasma dielectric tensor. The computational domain and typical configuration of periodic magnetic field of $B(z)=1.2+0.6\cos(2\pi z/3.63)$~T with $9$ magnetic mirrors and absorbing beach of length $7.74$~m are shown in Fig.~\ref{fg1}. 
To vary the number of mirrors ($N=1-10$) , the periodic area will be increased or decreased from the left end while the depth remains unchanged ($M=0.6/1.2$), and to vary the depth of mirrors ($M=0.1/1.2-1.0/1.2$), the number is kept to $N=9$. Other parameters include plasma density $n_e=9.2\times10^{19}\exp(-10.2~r^2)~\textrm{m}^{-3}$, electron temperature $8$~eV, helium ion species, blade antenna of length $L_a=1$~m and radius $R_a=0.005$~m, and chamber radius of $0.5$~m. Please note that the field strength has been increased by $10$ times to lower the collisional damping rate of Alfv\'{e}n waves (according to $3\omega^2\nu_{ei}/(2\omega_{ci}|\omega_{ce}|)$\cite{Braginskii:1965aa, Chang:2014aa}), and the plasma density is simutaneously increased by $100$ times to maintain the same phase velocity of Alfv\'{e}n waves for comparison with the previous experiment (according to $v_A=B/\sqrt{\mu_0 n_i m_i}$).

\begin{figure}[ht]
\begin{center}$
\begin{array}{ll}
(a)&(b)\\
\includegraphics[width=0.49\textwidth,angle=0]{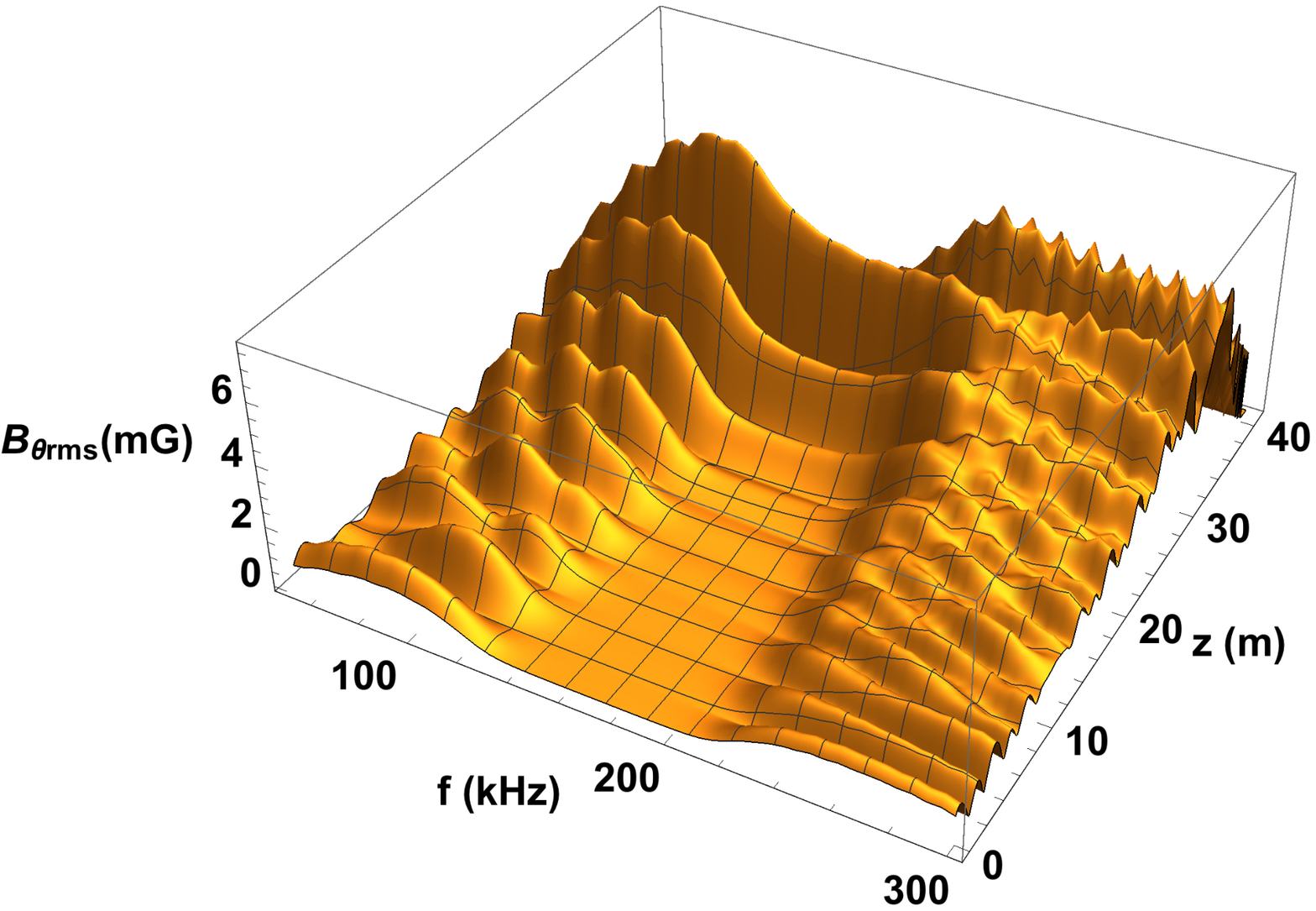}&\includegraphics[width=0.495\textwidth,angle=0]{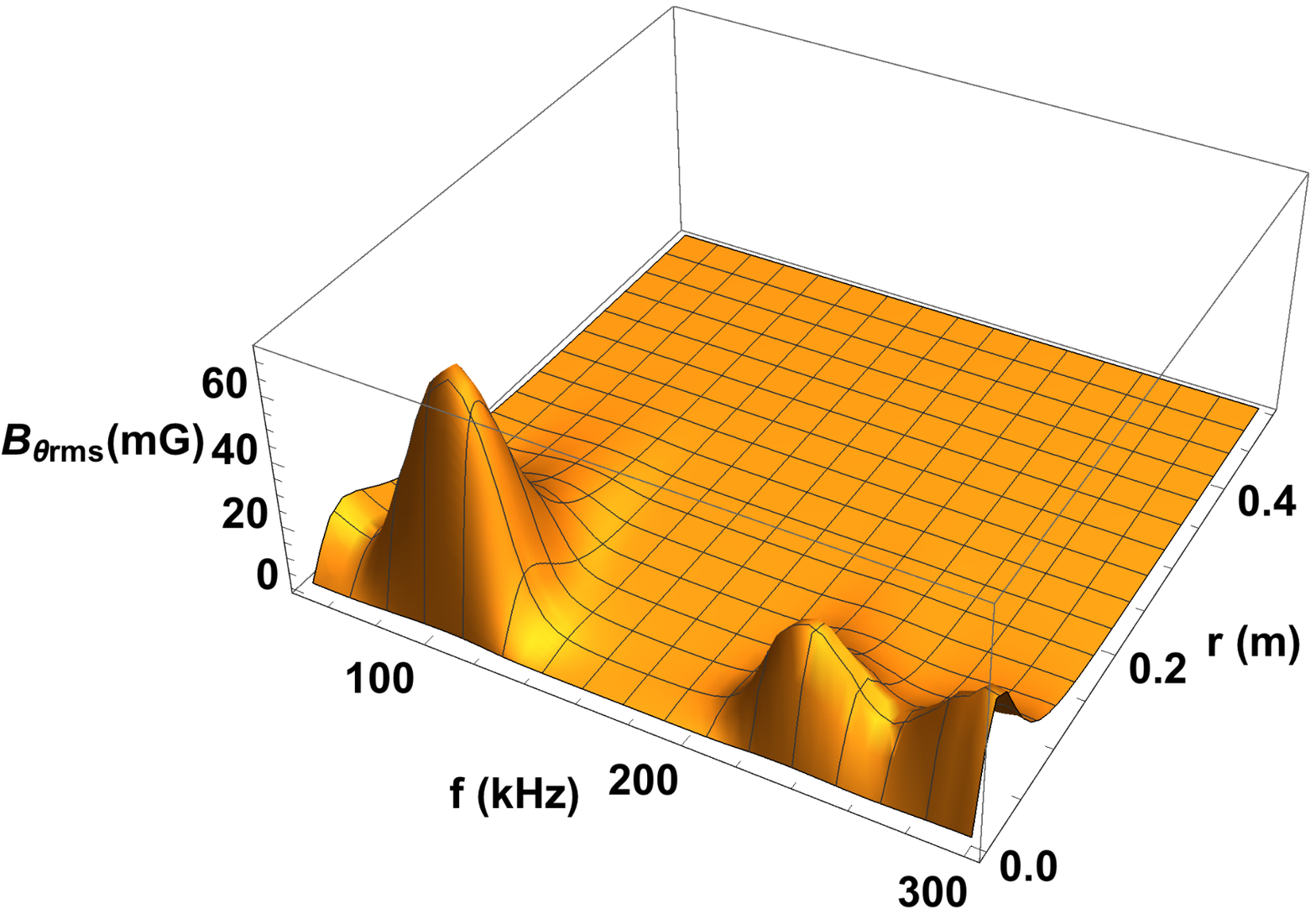}
\end{array}$
\end{center}
\caption{Variations of spectral gap in the axial ($r=0$~m) (a) and radial ($z=3.98$~m) (b) directions for magnetic field of $B(z)=1.2+0.6\cos(2\pi z/3.63)$~T with $9$ magnetic mirrors (Fig.~\ref{fg1}). The analytical center of spectral gap is $f_{B}=188$~kHz on axis. }
\label{fg2}
\end{figure}

\begin{figure}[ht]
\begin{center}
\includegraphics[width=0.5\textwidth,angle=0]{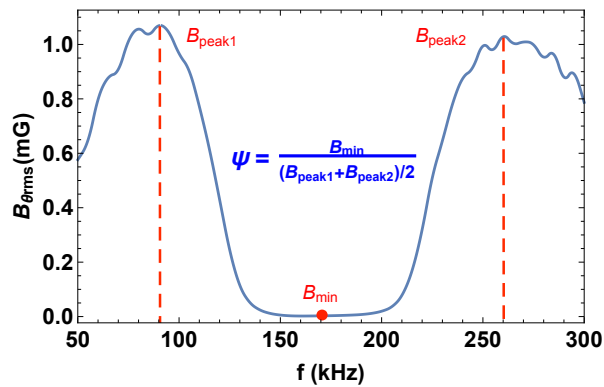}
\end{center}
\caption{Typical spectral gap measured at $z=0$~m and $r=0$~m for magnetic field of $B(z)=1.2+0.6\cos(2\pi z/3.63)$~T with $9$ magnetic mirrors (Fig.~\ref{fg1}), illustrating the definition of bottom-edge ratio: $\psi=B_{min}/[(B_{peak1}+B_{peak2})/2]$.}
\label{fg3}
\end{figure}

\begin{figure}[ht]
\begin{center}$
\begin{array}{ll}
(a)&(b)\\
\includegraphics[width=0.49\textwidth,angle=0]{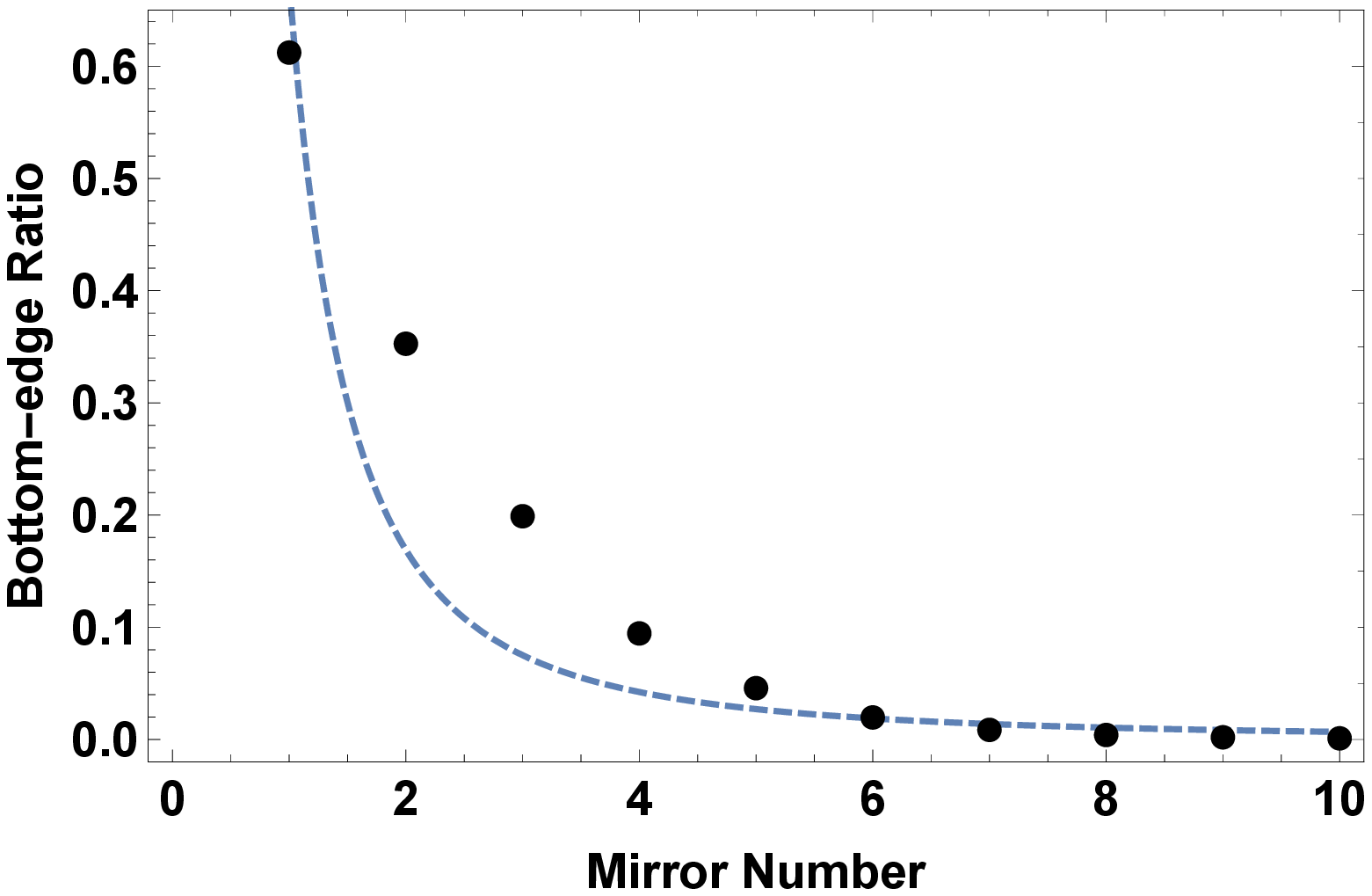}&\includegraphics[width=0.49\textwidth,angle=0]{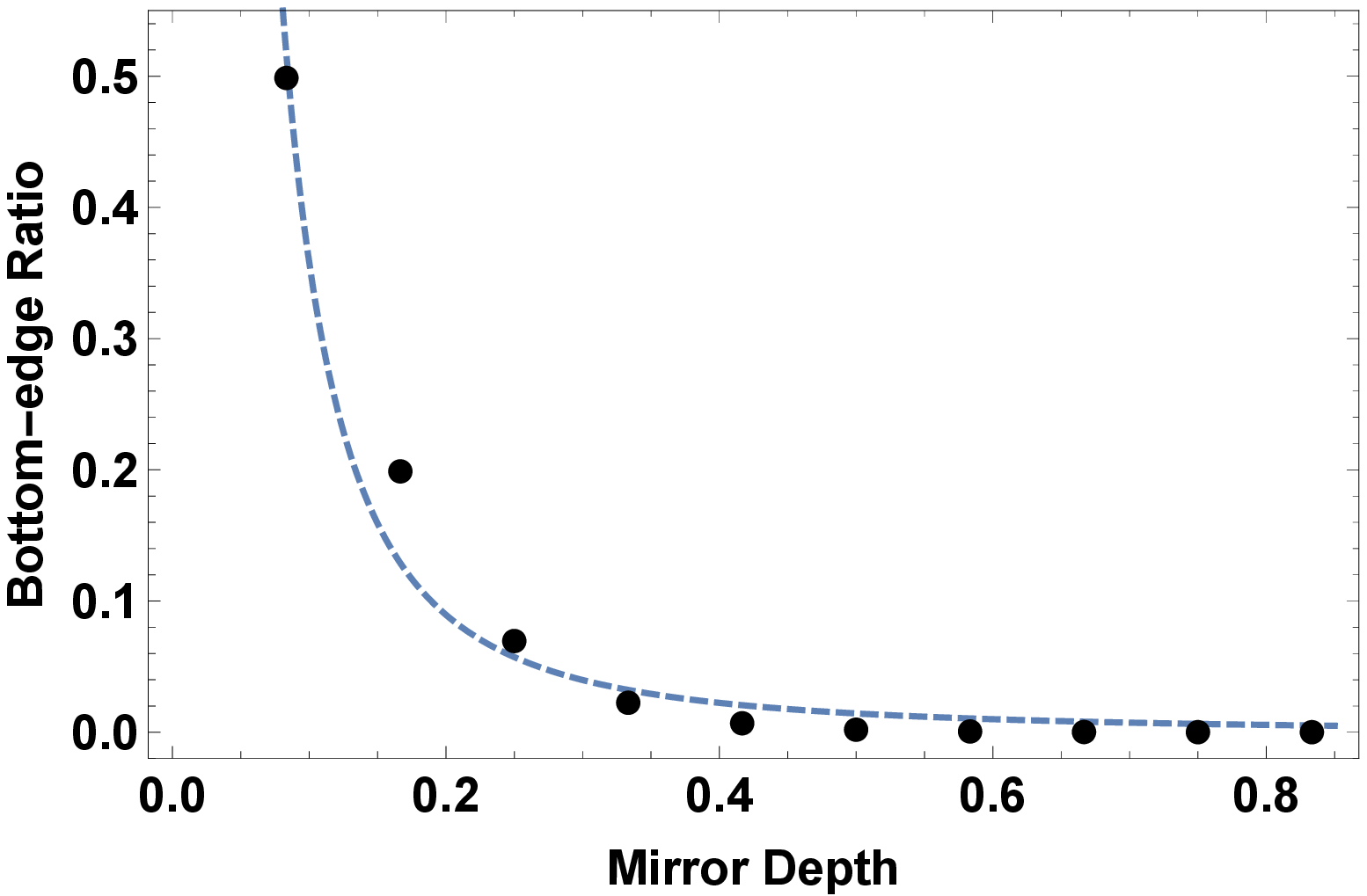}
\end{array}$
\end{center}
\caption{The dependence of bottom-edge ratio of spectral gaps on the number (a) and depth (b) of periodic modulations. The fitted dashed lines have analytical forms of $\psi=0.6756 N^{-2}$ and $\psi=0.0036 M^{-2}$ respectively.}
\label{fg4}
\end{figure}

The computed spectral gap for $9$ mirrors and depth of $0.5$ is given in Fig.~\ref{fg2}, which is clear in the whole axial and radial directions. The smooth radial profile of wave magnetic field also implies the absence of continuum damping resonance. The center frequency of spectral gap agrees resonably with the analytical estimate of $f_B=v_A/2L=188$~kHz on axis.
To quantify how well a spectral gap is formed and characterize the ``noise" level inside, a new parameter of bottom-edge ratio is introduced. Taking a slice of spectral gap at $z=0$~m from Fig.~\ref{fg2}(a) for example, the bottom-edge ratio $\psi$ is defined to be $\psi=B_{min}/[(B_{peak1}+B_{peak2})/2]$, as illustrated in Fig.~\ref{fg3}, namely the ratio of minimum bottom field to averaged peak field of lower and upper edges. 
The variations of $\psi$ with the number and depth of magnetic mirrors are shown in Fig.~\ref{fg4}. Here, all spectral gaps chosen to calculate $\psi$ are taken from the left end of plasma cylinder, namely $z=0$~m. The fitted dashed lines are in analytical forms of $\psi=0.6756 N^{-2}$ and $\psi=0.0036 M^{-2}$ respectively. 
It can be seen that the bottom-edge ratio scales inversely with the square of the number and depth of periodic modulations, and the ratio is very small (less than $0.01$) when $N\geq 7$ and $M\geq 0.4$, implying the formation of clear spectral gaps. This explains the previously unsuccessful formation of clear and global spectral gap on the LAPD (LArge Plasma Device\cite{Gekelman:1991aa}) where the employed number and depth of magnetic mirrors are $N=4.5$ at most and $M=0.38$, respectively.\cite{Zhang:2008aa} Interestingly, the scaling of $\psi\propto N^{-2}$ is consistent with an analytical work from optics, which studies the scattering of a probe laser beam going through periodic vacuum polarizations and states that the enhanced interference due to Bragg's reflection are proportional to $N^2$.\cite{Kryuchkyan:2011aa} More periods result in stronger Bragg's interference and thereby clearer spectral gap which is less ``noisy". The scaling of $\psi\propto M^{-2}$, which was not reported before, agrees better with computations than the scaling of $\psi\propto N^{-2}$ does. 

\begin{figure}[ht]
\begin{center}$
\begin{array}{ll}
(a)&(b)\\
\includegraphics[width=0.49\textwidth,angle=0]{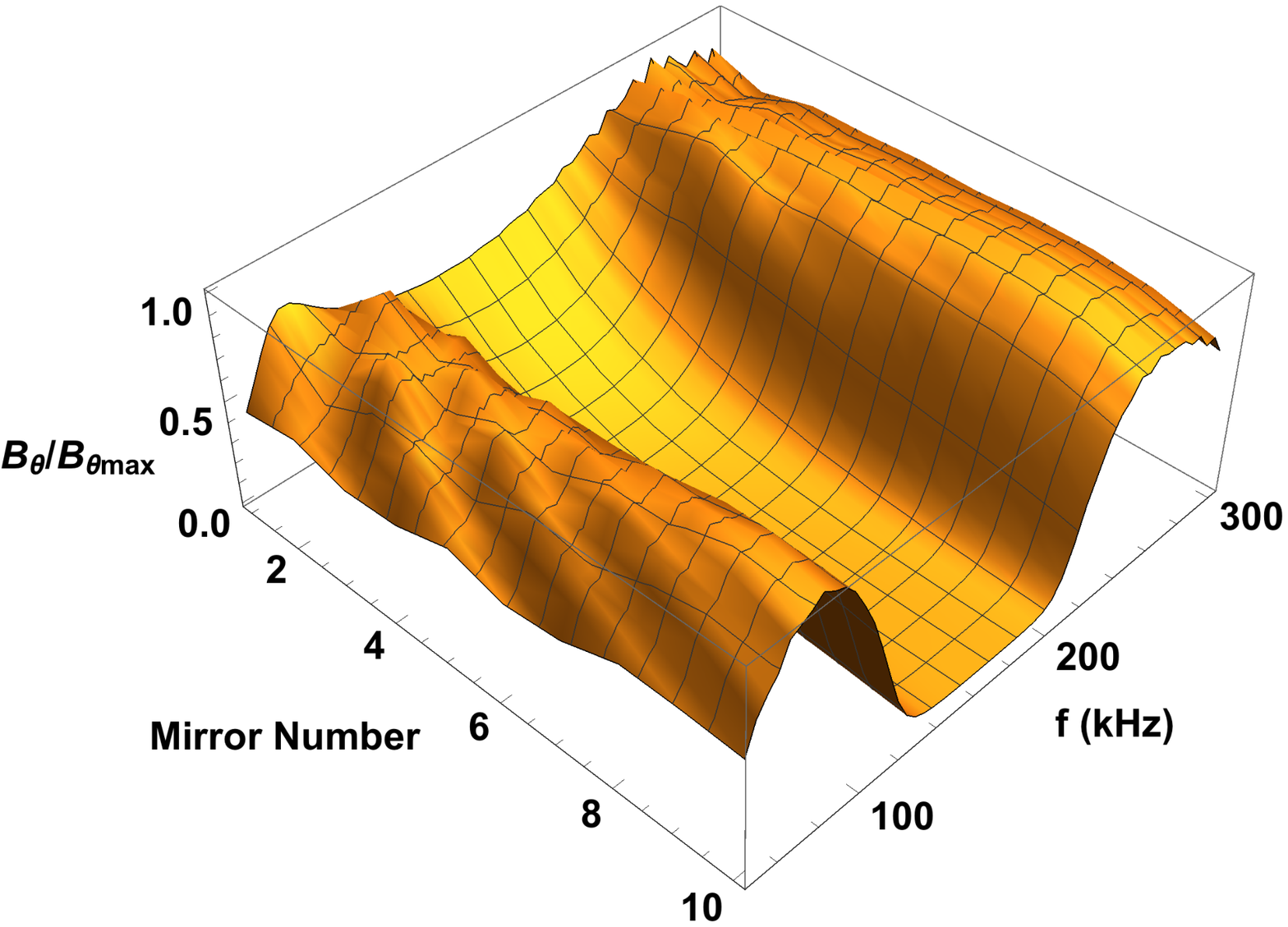}&\includegraphics[width=0.495\textwidth,angle=0]{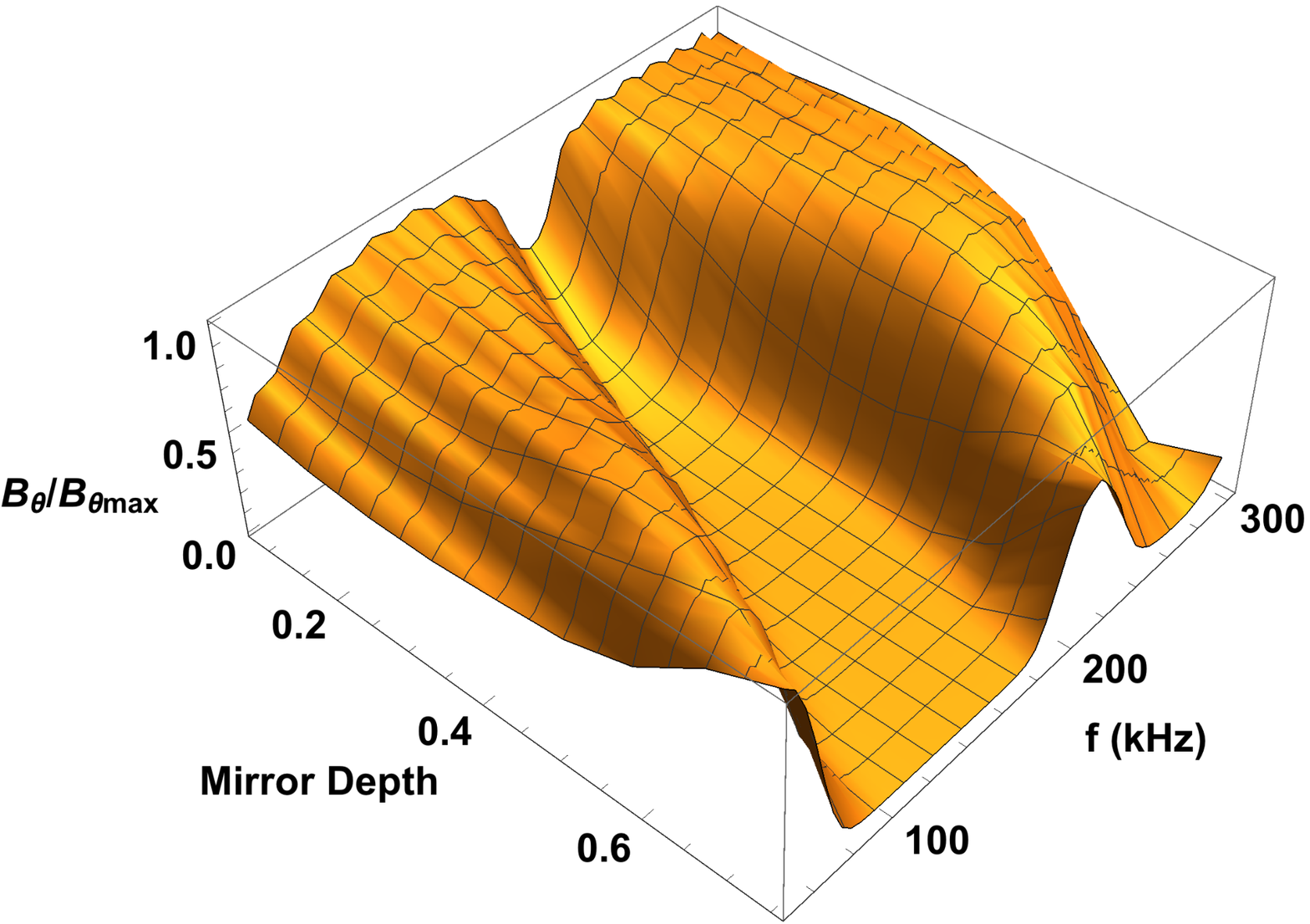}
\end{array}$
\end{center}
\caption{Surface plots of normalized wave magnetic field ($z=0$~m, $r=0$~m) as a function of driving frequency and: (a) mirror number, (b) mirror depth.}
\label{fg5}
\end{figure}

\begin{figure}[ht]
\begin{center}$
\begin{array}{ll}
(a)&(b)\\
\includegraphics[width=0.49\textwidth,height=0.4\textwidth,angle=0]{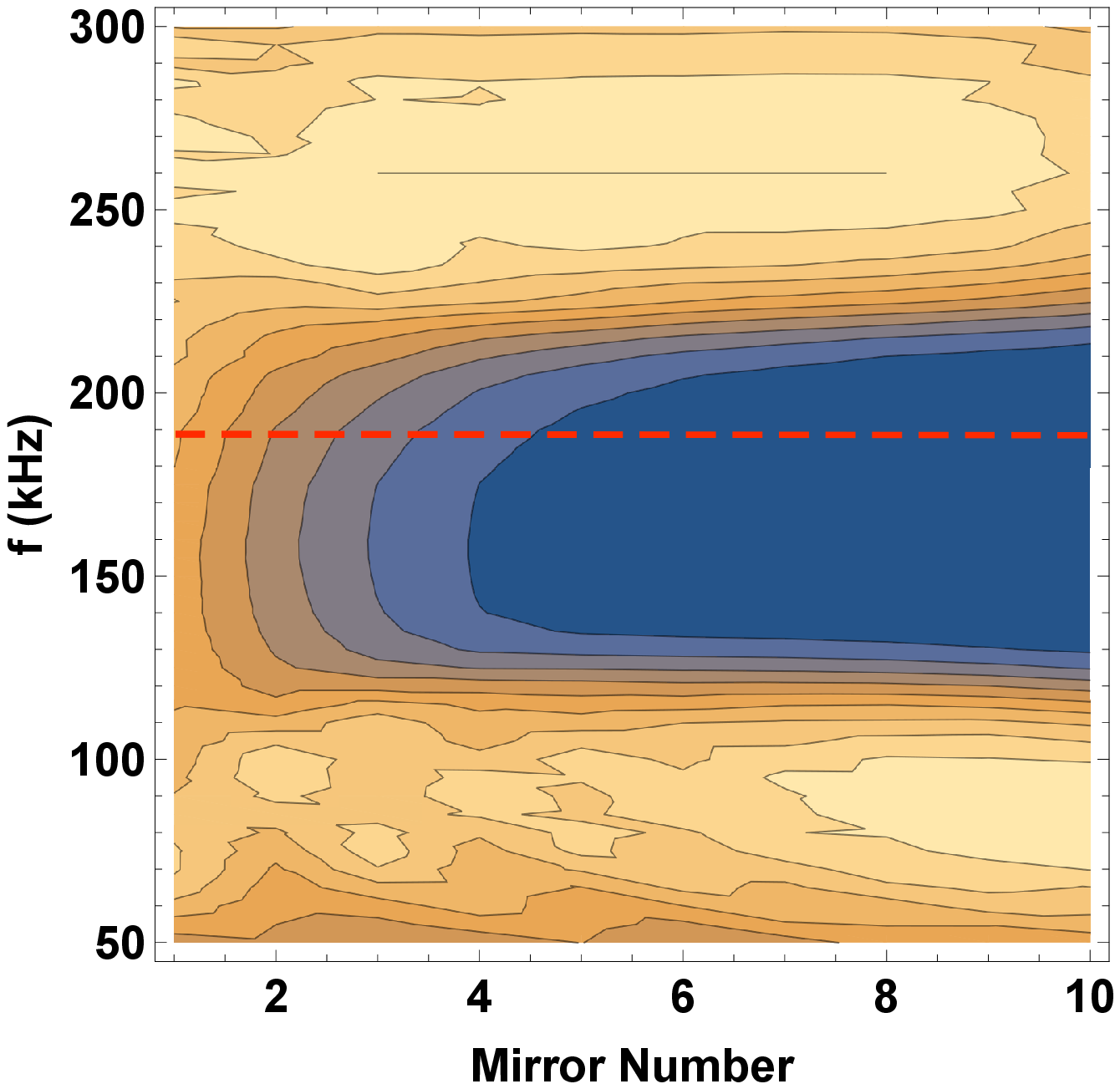}&\includegraphics[width=0.49\textwidth,height=0.4\textwidth,angle=0]{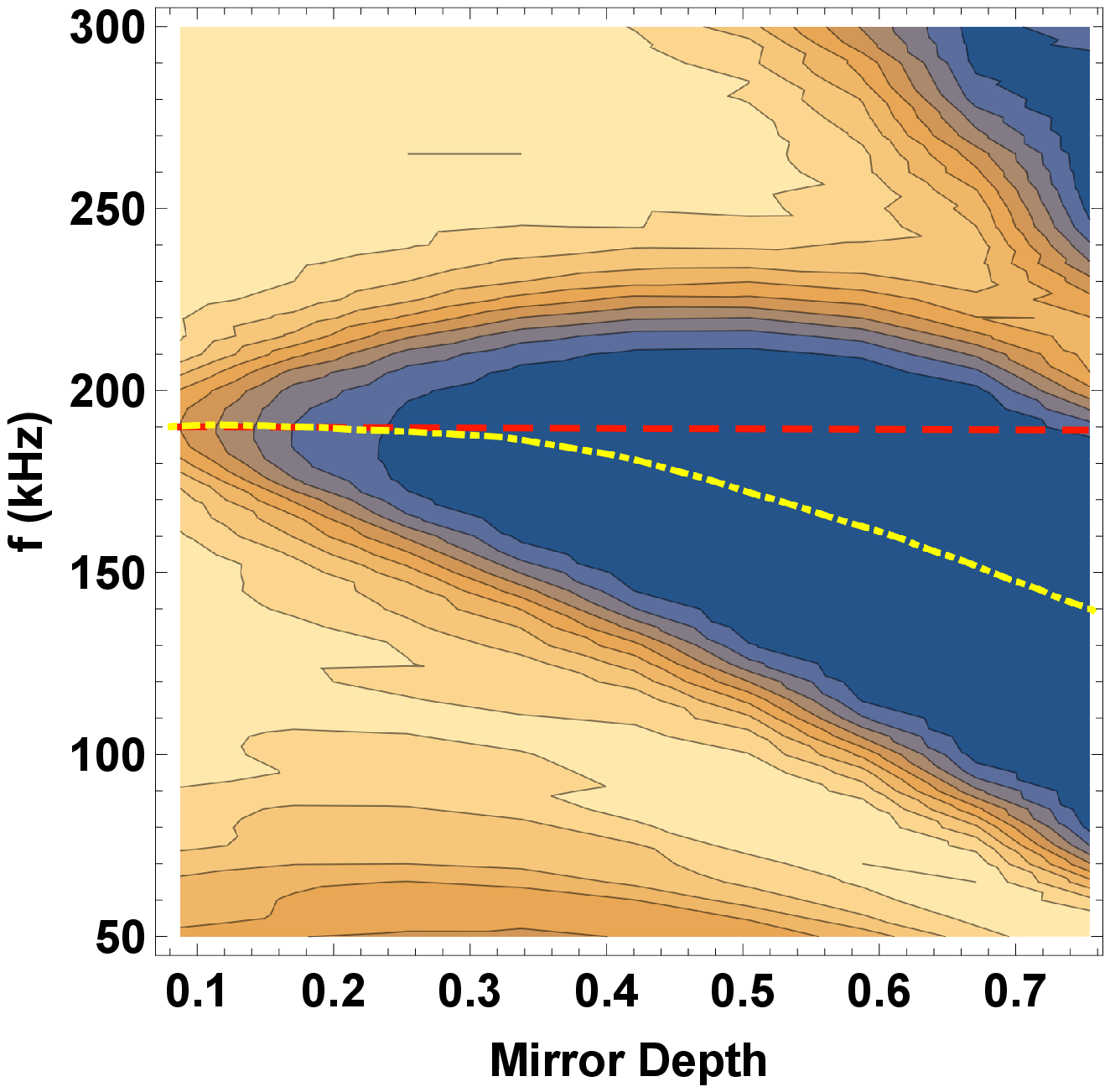}
\end{array}$
\end{center}
\caption{Contour plots of Fig.~\ref{fg5} showing better agreement between analytical center frequency of $f_B=188$~kHz (red dashed line) and computations for big $N$ (a) and small $M$ (b). The center of computed spectral gap (yellow dot-dashed line drawn by hand) also shows a parabolic shape of descending frequency shift when the depth is increased.}
\label{fg6}
\end{figure}

To show a full picture of the dependence of spectral gap on the number and depth of periodic modulations, variations of normalized wave magnetic field at $z=0$~m and $r=0$~m are plotted in Fig.~\ref{fg5} in the space of $(f, N)$ and $(f, M)$, respectively. It is obvious that spectral gap becomes more clear when there are sufficient magnetic mirrors and when the mirror depth is big enough. 
Contour plots of this dependence are given in Fig.~\ref{fg6}. They show that the analytical estimate of center frequency agrees with computations better for big $N$ and small $M$, as usually assumed in theoretical analysis. More interestingly, as the depth is increased, the center of computed spectral gap show a parabolic shape of descending frequency shift. Although similar frequency shift was observed on Alfv\'{e}nic gap eigenmode for increased depth,\cite{Chang:2016aa} this parabolic shape of frequency shift on spectral gap has not yet been revealed and is believed to be a new discovery. The underlying physics may be associated with squeezed plasma configuration and particle reflection from mirror throat for big modulation depth, which is true in reality but not considered in the present work due to the limitations of EMS. 

In summary, this letter introduces a new parameter to characterize how well a spectral gap is formed, namely bottom-edge ratio (the ratio of minimum bottom field to the averaged peak field at lower and upper edges), and shows that this ratio scales inversely with the square of the number and depth of periodic modulations. For the parameters of LAPD, the minimum number and depth of magnetic mirrors are $7$ and $0.4$, respectively, for clear spectral gap formation, which is of great practical interest for future experiments. The center of computed spectral gap shows a parabolic shape of descending frequency shift when the modulation depth is increased. The underlying physics may be related to plasma squeezing effect and particle reflection from magnetic mirrors which, however, is not considered in EMS and left for future research. 

This work is supported by various fundings: National Natural Science Foundation of China (11405271), China Postdoctoral Science Foundation (2017M612901), Chongqing Science and Technology Commission (cstc2017jcyjAX0047), Chongqing Postdoctoral Special Foundation (Xm2017109), Fundamental Research Funds for Central Universities (YJ201796), Pre-research of Key Laboratory Fund for Equipment (61422070306), and Laboratory of Advanced Space Propulsion (LabASP-2017-10). 

\section*{References}
\bibliographystyle{unsrt}

\end{document}